\documentclass[aps,reprint,pre,showpacs,superscriptaddress]{revtex4-2}
\usepackage{amsmath}
\usepackage{amssymb}
\usepackage{graphicx}
\usepackage{color}
\usepackage{bm}
\usepackage{float}
\usepackage{siunitx}
\usepackage{mathtools}
\usepackage{MnSymbol}
\usepackage{dcolumn} 

\usepackage{babel}

\definecolor{linkcolor}{rgb}{0.56640625,0.3359375,0.42578125} 

\usepackage[pdftex,colorlinks=true,
pdfstartview=FitV,
linkcolor= linkcolor,
citecolor= linkcolor,
urlcolor= linkcolor,
hyperindex=true,
hyperfigures=false]
{hyperref} 

\begin{document}

\title{Avalanches in the Random Organization Model with long-range interactions}

\author{Tristan Jocteur}
\author{Kirsten Martens}
\author{Romain Mari}
\author{Eric Bertin}

\affiliation{Univ.~Grenoble-Alpes, CNRS, LIPhy, 38000 Grenoble, France}

\begin{abstract}
Oscillatory sheared suspensions, when observed stroboscopically, exhibit a reversible-irreversible transition as a function of the strain amplitude, which is a kind of absorbing phase transition.
So far studies of this transition focused on global quantities, e.g. quantifying the irreversibility on one side of the transition or the time to reach a reversible state on the other side. Here, motivated by the kin depinning transition, we focus on the intermittent  dynamics near the transition.
We perform simulations of a modified Random Organization Model (ROM), a minimal particle model which we recently adapted to take into account the generic presence of long-range interactions mediated by the fluid,
taking the power-law-decay exponent $\alpha$ as an additional control parameter of the model.
We show that at the absorbing phase transition, this model displays power-law-distributed avalanches.
We characterize the avalanche statistics in terms of avalanche size, duration and number of particles involved, and
we determine the associated exponents.
By varying the exponent $\alpha$, the fractal dimension of avalanches crosses space dimension $d$, inducing a qualitative change of the
spatial structure of avalanches, from compact avalanches when interactions have a short range, to sparse avalanches when interactions are long-ranged.
Finally, we characterize the clusters within the avalanches, which we also find power-law distributed.
\end{abstract}

\maketitle

\section{Introduction}

Intermittent dynamics in the form of avalanches is a common feature of material deformation. Across a wide range of phenomena including crystal plasticity, yielding of amorphous solids, flow transition of yield stress fluids, granular flow and depinning, slow external driving gives rise to sudden, collective rearrangements whose sizes, durations, and spatial extents are broadly distributed~\cite{zapperiDynamicsFerromagneticDomain1998,miguelIntermittentDislocationFlow2001,rosso_avalanche-size_2009,talamaliAvalanchesPrecursorsFinitesize2011,wieseTheoryExperimentsDisordered2022}. 
Avalanche statistics provide a powerful dynamical characterization of criticality, often revealing scaling laws and geometric properties that go beyond those accessible from steady-state observables alone.

In particle suspensions subject to oscillatory shear, a closely related phenomenon occurs in the form of a reversible-irreversible transition \cite{pineChaosThresholdIrreversibility2005,corteRandomOrganizationPeriodically2008,corteSelfOrganizedCriticalitySheared2009,guastoHydrodynamicIrreversibilityParticle2010,geRheologyPeriodicallySheared2022}.
Below a critical driving amplitude after a transient dynamics during which they experience collisions, particles organize into following strictly periodic motion, defining an absorbing state.
Above this threshold, irreversible rearrangements due to collisions persist indefinitely, leading to diffusive dynamics and sustained activity.
The reversible-irreversible transition has been observed experimentally in non-Brownian suspensions \cite{pineChaosThresholdIrreversibility2005,corteRandomOrganizationPeriodically2008,corteSelfOrganizedCriticalitySheared2009,guastoHydrodynamicIrreversibilityParticle2010,geRheologyPeriodicallySheared2022} as well as different types of complex fluids~\cite{reichhardtReversibleIrreversibleTransitions2023}, like foams~\cite{mukherjiStrengthMechanicalMemories2019}, microemulsions~\cite{jeanneretGeometricallyProtectedReversibility2014,weijsEmergentHyperuniformityPeriodically2015}, soft glasses~\cite{fioccoOscillatoryAthermalQuasistatic2013,keimMechanicalMicroscopicProperties2014,himanagamanasaExperimentalSignaturesNonequilibrium2014,regevReversibilityCriticalityAmorphous2015,ederaYieldingMicroscopeMultiscale2024}, or dry granular materials~\cite{royerPreciselyCyclicSand2015}.
This transition has been interpreted~\cite{corteRandomOrganizationPeriodically2008,corteSelfOrganizedCriticalitySheared2009,menonUniversalityClassReversibleirreversible2009,tjhungHyperuniformDensityFluctuations2015,nessAbsorbingStateTransitionsGranular2020,mariAbsorbingPhaseTransitions2022,jocteur2025EPJE} as a nonequilibrium absorbing phase transition~\cite{lubeckUniversalScalingBehavior2004}.
To date, however, studies of this transition have primarily focused on steady-state properties such as diffusion, activity, and hyperuniformity~\cite{schrenkCommunicationEvidenceNonergodicity2015,hexnerHyperuniformityCriticalAbsorbing2015,hexnerNoiseDiffusionHyperuniformity2017}.
Dynamical properties have received comparatively little attention besides characterization of the critical slowing down~\cite{corteRandomOrganizationPeriodically2008,tjhungHyperuniformDensityFluctuations2015}.


The Random Organization Model (ROM) \cite{corteRandomOrganizationPeriodically2008,hexnerHyperuniformityCriticalAbsorbing2015,hexnerNoiseDiffusionHyperuniformity2017,tjhungHyperuniformDensityFluctuations2015,schrenkCommunicationEvidenceNonergodicity2015} provides a minimal theoretical framework for the reversible-irreversible transition in suspensions.
In this stroboscopic model, particles involved in overlaps at the end of a shear cycle are randomly displaced, while non-overlapping particles remain immobile.
Iterating this rule leads either to an absorbing state with no active particles or to an active steady state, depending on control parameters such as particle density or driving amplitude.
The ROM falls into the conserved directed percolation class (also called Manna class)~\cite{menonUniversalityClassReversibleirreversible2009}, which itself is equivalent to the short-range depinning class~\cite{ledoussalExactMappingStochastic2015}.
Near the depinning transition, activity proceeds an underlying avalanche dynamics~\cite{rosso_avalanche-size_2009}, suggesting that avalanches also occur near the RIT.
Despite extensive work on the ROM and related models \cite{corteRandomOrganizationPeriodically2008,hexnerHyperuniformityCriticalAbsorbing2015,hexnerNoiseDiffusionHyperuniformity2017,tjhungHyperuniformDensityFluctuations2015,schrenkCommunicationEvidenceNonergodicity2015,tjhungCriticalityCorrelatedDynamics2016,mariAbsorbingPhaseTransitions2022,jocteur2025EPJE}, a systematic characterization of avalanches at the reversible-irreversible transition in suspensions is still lacking.
So far, only a characterisation of the size and duration exponents has been performed in a modified model including sedimentation~\cite{corteSelfOrganizedCriticalitySheared2009}.

In real suspensions, particle rearrangements are not purely local.
Hydrodynamic interactions mediate long-range displacements through the surrounding fluid, transmitting activity over distances much larger than the particle size \cite{weijsEmergentHyperuniformityPeriodically2015}.
Recent numerical work has shown that incorporating long-range mediated interactions into random organization models profoundly affects the nature of the absorbing transition, leading to regimes with vanishing critical fluctuations and departures from standard conserved directed percolation behavior \cite{mariAbsorbingPhaseTransitions2022,jocteur2025EPJE}.
Similarly, yielding transitions in amorphous solids have been reinterpreted as absorbing phase transitions with long-range interactions and zero modes, forming a continuum of universality classes controlled by interaction range \cite{jocteurYieldingAbsorbingPhase2024a}.
These developments highlight the central role of long-range interactions in shaping nonequilibrium critical behavior.


In this work, we provide a systematic study of avalanche dynamics at the reversible-irreversible transition in suspensions.
Following Ref.~\cite{jocteur2025EPJE}, we consider a mediated random organization model in which local rearrangements generate long-range displacements through a power-law interaction kernel decaying as $r^{-\alpha}$.
By tuning the exponent $\alpha$, we continuously interpolate between effectively short-range and strongly long-range interactions while preserving the absorbing-state structure of the transition.
Focusing on criticality, we characterize avalanches in terms of their size, duration, number of participating particles, and spatial extent.

We find that avalanches remain scale-free at the transition but exhibit continuously varying exponents as the interaction range is tuned.
The fractal dimension of avalanches evolves with $\alpha$ and crosses the spatial dimension at a characteristic interaction range, signaling a qualitative change in avalanche geometry.
We also briefly discuss the validity of scaling relations between avalanche exponents, as well as some statistical properties of clusters found within avalanches.
Our results establish avalanche dynamics as a sensitive probe of interaction-range effects at absorbing phase transitions and open the way to direct comparisons with experiments on sheared suspensions.


\section{Model}

Close to the critical point, the behavior of cyclically sheared suspensions may be modeled as an extension of the ROM, taking into account long-range hydrodynamic interactions \cite{mariAbsorbingPhaseTransitions2022,jocteur2025EPJE}. We briefly recall here the definition of the original version of the ROM, before introducing its generalization with long-range mediated interactions.

\subsection{Random Organization Model}

In its original and minimal version, the ROM~\cite{corteRandomOrganizationPeriodically2008,hexnerHyperuniformityCriticalAbsorbing2015,hexnerNoiseDiffusionHyperuniformity2017,tjhungHyperuniformDensityFluctuations2015,schrenkCommunicationEvidenceNonergodicity2015}
describes the suspension dynamics in a stroboscopic way, considering only short-range interactions.
We focus here on the two-dimensional version of the model.
The model is composed of $N_d$ discs of diameter $D=1$ and position $\bm{r}^t_i$ ($i=1,\dots,N_d$) at time $t$,
in a square box of linear extension $L$, with periodic boundary conditions.
When the distance $r^t_{ij} = |\bm{r}^t_i -\bm{r}^t_j|$ between particles $i$ and $j$ satisfies $r^t_{ij}<D$ (i.e., particles overlap), the corresponding particles are said to be active; other particles are then called passive. 
Physically, active particles can be interpreted as particles making a contact with another particle during a cycle in an oscillatory shear experiment.
The diameter $D$ may thus be seen as an effective diameter implicitly accounting for shear amplitude, rather than the hard-core diameter of suspended particles.
The discrete-time dynamics of the ROM goes as follows.
At a given time $t$, all active particles are randomly moved as
\begin{equation}
    \bm{r}^{t+1}_i = \bm{r}^{t}_i + \bm{\delta}_{a,i}\, ,
\end{equation}
where the random displacement $\bm{\delta}_{a,i}$ is drawn from a probability distribution of characteristic size $\Delta_a = D$.
These random kicks model the displacements resulting from contacts between particles when cyclically shearing a suspension.
The active or passive nature of the particles is updated from the positions reached at the end of the time step.
In case no more active particles are present, the dynamics stops and the system falls into an absorbing state.

The key control parameter in the model is the particle area fraction, $\phi = N_d \pi (D/2)^2/L^2$.
The statistically observed behavior in the long time limit crucially depends on whether $\phi$ is smaller or larger than a critical value $\phi_\mathrm{c}$.
When $\phi<\phi_\mathrm{c}$, an absorbing state with no active particles is always eventually observed.
In contrast, when $\phi > \phi_\mathrm{c}$, a stationary state is reached, and the fraction $A$ of active particles fluctuates around its mean value $\langle A \rangle>0$. 
The ROM thus displays an absorbing phase transition characterized by the order parameter $\langle A \rangle$, and controlled by the area fraction $\phi$.

\subsection{ROM with mediated interactions ($\alpha$-ROM)}

As immediately seen from its definition, the ROM neglects long-range hydrodynamic interactions, while such interactions may strongly impact the critical properties of the model
\cite{weijsEmergentHyperuniformityPeriodically2015,mariAbsorbingPhaseTransitions2022,jocteur2025EPJE}.
In the stroboscopic dynamics of sheared suspensions, the presence of hydrodynamic interactions mediated by the suspending fluid leads to small displacements of passive particles at each cycle, due to the irreversible contact dynamics of active particles (see \cite{jocteur2025EPJE} for a more detailed discussion of the underlying physical mechanisms and of modeling choices).
Here, following \cite{jocteur2025EPJE}, we assume that active particles induce random displacements of passive particles located at a distance $r=|\mathbf{r}|$, with a typical displacement of passive particles given by a physical propagator $\mathcal{G}(\mathbf{r}) \sim 1/r^{\alpha}$, with $\alpha>0$ the exponent characterizing long-range interactions.
We call $\alpha$-ROM this generalized model, to emphasize the key role played by the power-law decay exponent $\alpha$ of mediated interactions.
The random displacements induced by several active particles on the same passive particles simply add up, and are assumed to be uncorrelated, so that their variance is additive.
In practice, and for the sake of numerical efficiency, we implement a coarse-grained long-range interaction kernel, rather than explicitly simulating all particle pair interactions.
The slight discrepancy with respect to the exact particle-pair-interaction evaluation is expected to be irrelevant as far as critical properties are concerned.
To perform the coarse-graining, space is divided into a regular square grid of boxes of linear size $D$, and an activity field $A$ is evaluated in each box 
$\mathfrak{b}$ centered on position $\bm{r}_\mathfrak{b}$ as $A^t_\mathfrak{b} = \mathcal{N}_\mathfrak{b}^a /D^2$, with $\mathcal{N}_\mathfrak{b}^a$ the number of active particles whose center is located within box $\mathfrak{b}$.

Any passive particle $i$, located in a box $\mathfrak{b}$, is moved at a given time step by a random displacement vector $\bm{\delta}_{\mathrm{p},i}$.
Each component of the vector $\bm{\delta}_{\mathrm{p},i}$ is drawn from a Gaussian distribution with zero mean, and variance $\Delta_{\mathrm{p},i}^2$ determined by
\begin{equation}
    \Delta_{\mathrm{p},i}^2 = \sum_{\mathfrak{b}'} G(r_{\mathfrak{b}'\mathfrak{b}}) A_{\mathfrak{b}'}
    \label{eq:passive_step_size}
\end{equation}
where $r_{\mathfrak{b}'\mathfrak{b}}$ is the center-to-center distance between boxes $\mathfrak{b}'$ and $\mathfrak{b}$.

Physically, the propagator $G(r)$ appearing in Eq.~\eqref{eq:passive_step_size} may be interpreted as the square of the physical propagator $\mathcal{G}(\mathbf{r})$ mentioned earlier, so that $G(r) \sim 1/r^{2\alpha}$ at large distance $r$.
For practical purpose, it is convenient to choose the following form of the propagator $G(r)$,
\begin{equation}
    G(r) = \frac{c}{(1+r^2)^{\alpha}}
    \label{eq:kernel}
\end{equation}
since its discrete Fourier transform can be evaluated explicitly in two dimensions, which is useful to efficiently evaluate the discrete convolution appearing in Eq.~\eqref{eq:passive_step_size}.
The coefficient $c$ may depend on system size $L$ to ensure convergence of the convolution in $\eqref{eq:kernel}$ when $L\to \infty$.
For $\alpha >1$, we choose $c=0.25$, while for $\alpha <1$, we take $c = 0.25 L^{2\alpha-2}$ \cite{jocteur2025EPJE}. Note that the case $\alpha=1$ is singular and leads to logarithmic corrections which hinder a proper determination of the critical point, and we do not consider it here.

As limiting cases, the short-range behavior is recovered for  $\alpha\gtrsim 2$, while in the opposite limit of small $\alpha \lesssim 0.5$ values, a mean-field-like behavior is observed \cite{jocteur2025EPJE}. In between these asymptotic behaviors, critical exponents continuously depend the on long-range interaction exponent $\alpha$ \cite{jocteur2025EPJE}, in analogy with the critical behavior generically reported in the presence of long-range interactions at equilibrium~\cite{fisherCriticalExponentsLongRange1972} or for absorbing phase transitions~\cite{hinrichsenNonequilibriumPhaseTransitions2007}.

\section{Avalanche statistics}

\subsection{Definitions of avalanches and relevant observables}
\label{sec:avalanches_def}

At the reversible--irreversible transition, the dynamics at finite $N_d$ is active only during a finite time, as the system always eventually ends up in an absorbing configuration.
This system is however very sensitive to external perturbations, such as a small extra Brownian-like noise, that can reactivate the dynamics~\cite{hexnerEnhancedHyperuniformityRandom2017}.
In presence of a reactivation mechanism, the dynamics of the mediated ROM at the reversible--irreversible transition is then highly intermittent~\cite{hexnerEnhancedHyperuniformityRandom2017}.
Activity proceeds through bursts separated by quiescent periods during which the system resides in an absorbing configuration until a new reactivation occurs.
In practice, each burst needs to be initiated by a reactivation event (see below).
We refer to these bursts of activity as avalanches.
In the context of yielding, it has been demonstrated that avalanche statistics may depend on the reactivation protocol, but crucially the exponents associated with these avalanche statistics obey the same scaling relations irrespective of the reactivation protocol~\cite{jocteurProtocol2025}.

An avalanche is defined as a temporally connected sequence of active states, starting when activity becomes nonzero and ending when the system returns to an absorbing state.
An avalanche begins at the first cycle for which this number is nonzero and terminates at the first subsequent cycle for which activity vanishes.

We characterize avalanches through several complementary observables.
The avalanche \emph{duration} $T$ is defined as the number of consecutive cycles during which the number of active particles remains nonzero.
The avalanche \emph{size} $S$ is defined as the total number of active particle displacement events occurring during the avalanche,
\begin{equation}
S = \sum_{t=1}^{T} N_{\mathrm{act}}(t),
\end{equation}
where $N_{\mathrm{act}}(t)$ denotes the number of active particles at cycle $t$ (we recall that each active particle moves once per cycle).
In addition, we define the avalanche \emph{participation number} $N$ as the total number of distinct particles that become active at least once during the avalanche.

\subsection{Distribution of avalanche observables}

\begin{figure}[t]
    \centering
    \includegraphics[width=0.47\textwidth]{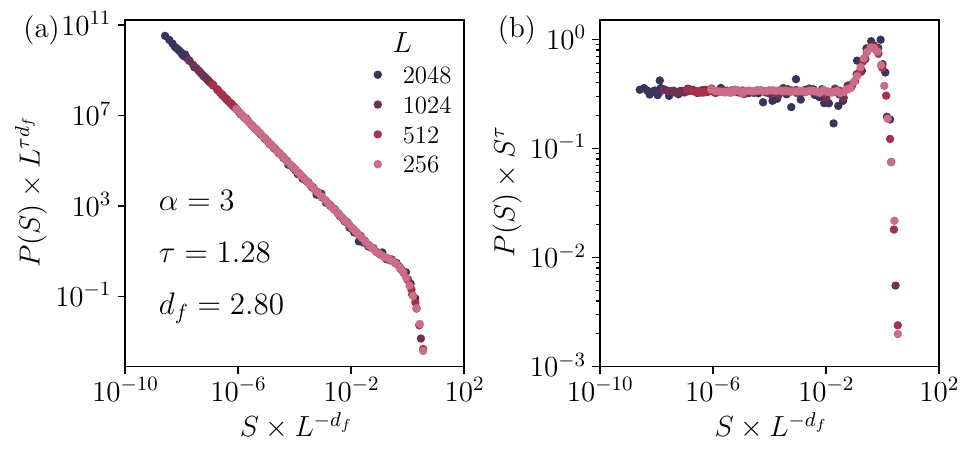}
    \includegraphics[width=0.47\textwidth]{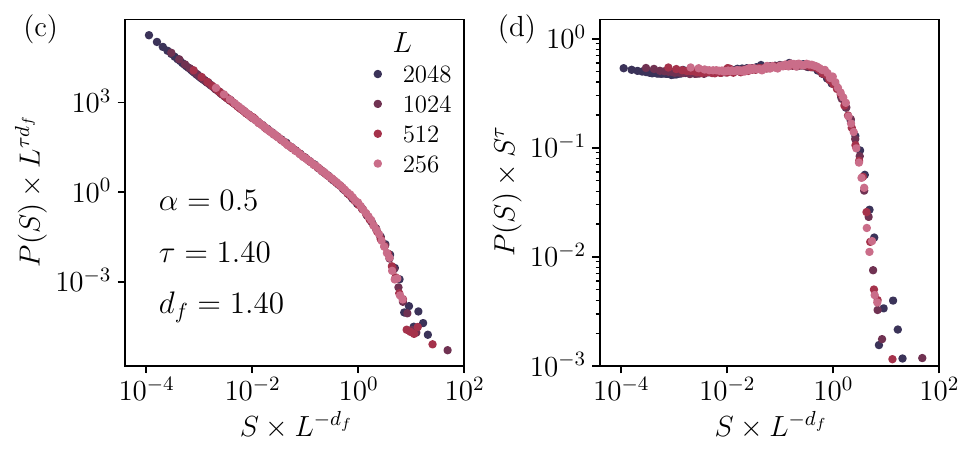}
    \caption{Avalanche size distribution $P(S)$ for $\alpha=3$ (a,b) and $\alpha=0.5$ (c,d). Left panels (a,c): $P(S)$ rescaled by system size $L$, showing a good data collapse on a power-law distribution $\propto S^{-\tau}$ followed by a size-dependent cutoff around $S_c \sim L^{d_f}$. Right panels (b,d): compensated plots obtained by dividing the distribution $P(S)$ by the estimated power law $S^{-\tau}$.}
    \label{fig:size}
\end{figure}

To characterize the avalanche statistics, we determine the distribution of avalanche size $S$, duration $T$ and number $N$ of involved particles. We respectively denote these distributions
as $P(S)$, $P(T)$ and $P(N)$.
We are interested in the critical properties of avalanches, and we thus set the packing fraction $\phi$ to its ($\alpha$-dependent) critical value $\phi_\mathrm{c}$.
To obtain a well-defined avalanche statistics, we  specify a reactivation protocol to trigger avalanches in a reproducible way.
 We first generate an absorbing state starting from a random initial configuration.
We then reactivate the dynamics  by turning a randomly chosen particle active. 
In experiment, reactivation can occur through sedimentation or Brownian motion~\cite{corteSelfOrganizedCriticalitySheared2009,keimGenericTransientMemory2011,paulsenMultipleTransientMemories2014}.
This process is then repeated each time the system falls into an absorbing state, until we reach a stationary state, that is, until convergence and stationarity of the avalanche statistics under repeated reactivations~\cite{hexnerEnhancedHyperuniformityRandom2017}.
Avalanche statistics are then sampled from this stationary dynamics
 for $\alpha \in \{ 3, 2, 1.75, 1.5, 1.25, 0.5 \}$ and for system sizes
$L \in \{ 256, 512, 1024, 2048 \}$. For the sake of clarity, we only display here avalanche distributions for the extreme values of the considered range of $\alpha$, namely $\alpha=3$ and $\alpha=0.5$.
Avalanche distributions obtained for $\alpha \in \{ 2, 1.75, 1.5, 1.25 \}$ are presented in Appendix~\ref{sec:app:avalanches} (see Figs.~\ref{fig:size:app}, \ref{fig:time:app} and \ref{fig:number:app}).

\begin{figure}[t]
    \centering
    \includegraphics[width=0.48\textwidth]{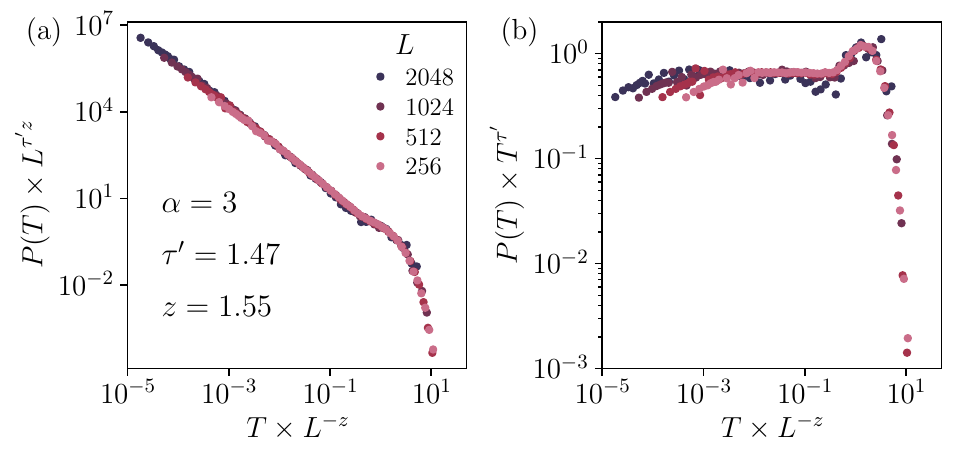}
    \includegraphics[width=0.48\textwidth]{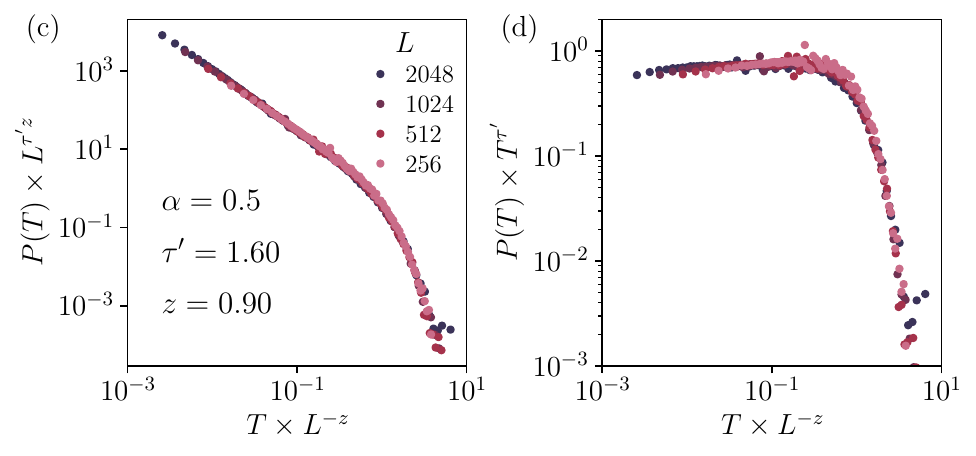}
    \caption{Avalanche duration distribution $P(T)$ for $\alpha=3$ (a,b) and $\alpha=0.5$ (c,d). Left panels (a,c): $P(T)$ rescaled by system size $L$, showing a good data collapse on a power-law distribution $\propto T^{-\tau'}$ followed by a size-dependent cutoff $T_c \sim L^z$. Right panels (b,d): compensated plots obtained by dividing the distribution $P(T)$ by the estimated power law $T^{-\tau'}$.}
    \label{fig:time}
\end{figure}

To investigate finite-size effects and clarify the behavior of the avalanche distributions $P(S)$, $P(T)$ and $P(N)$ with system size $L$, we perform a finite-size analysis of the distributions
$P(S)$, $P(T)$ and $P(N)$.
At criticality, we expect the probability distributions of avalanche observables to follow power-law forms, with prefactors and upper cutoffs depending on the system size $L$:

\begin{equation}
	\begin{aligned}
		&P(S) \sim S^{-\tau} g_S\left( \frac{S}{S_c} \right), \quad S_c \sim L^{d_f},\\
		&P(T) \sim T^{-\tau^\prime} g_T\left( \frac{T}{T_c} \right), \quad T_c \sim L^{z},\\
		&P(N) \sim N^{-\tau^{\prime\prime}} g_N\left( \frac{N}{N_c} \right), \quad N_c \sim L^{\chi},\\
	\end{aligned}
	\label{eq:AvDistribSusp}
\end{equation}
where the power-law exponents $\tau$, $\tau'$, $\tau''$ and the cutoff exponents $d_f$, $z$ and $\chi$ are to be determined from numerical data.
The exponent $d_f$ is called the fractal dimension of avalanches, and $z$ is called the dynamical exponent.
The functions $g_S(x)$, $g_T(x)$ and $g_N(x)$ are scaling functions going to a constant for $x\to 0$, and to zero for $x\to \infty$.

\begin{figure}[t]
    \centering
    \includegraphics[width=0.48\textwidth]{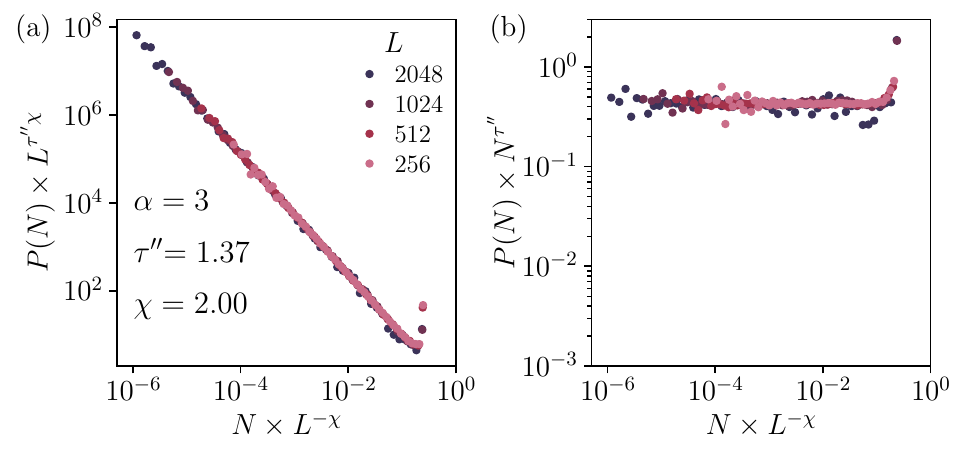}
    \includegraphics[width=0.48\textwidth]{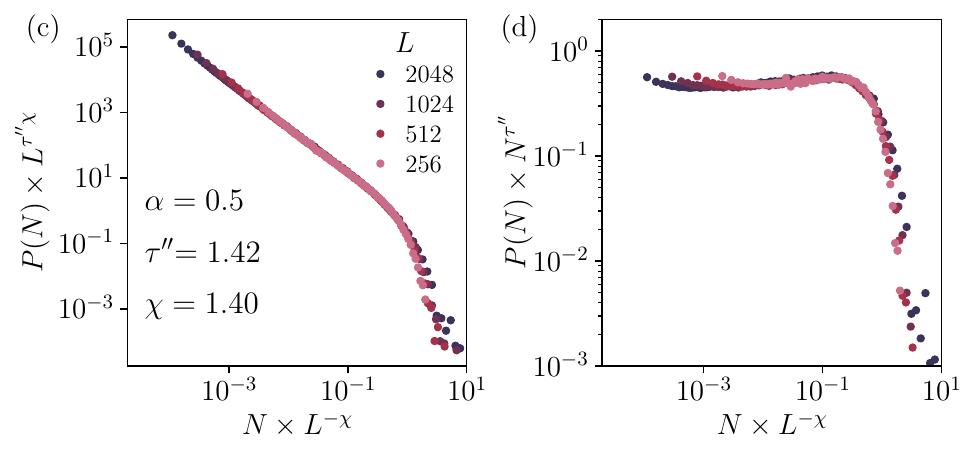}
    \caption{Distribution $P(N)$ of the number $N$ of particles involved in an avalanche, for $\alpha=3$ (a,b) and $\alpha=0.5$ (c,d). Left panels (a,c): $P(N)$ rescaled by system size $L$, showing a good data collapse on a power-law distribution $\propto N^{-\tau''}$ followed by a size-dependent cutoff $N_c \sim L^{\chi}$. Right panels (b,d): compensated plots obtained by dividing the distribution $P(N)$ by the estimated power law $N^{-\tau''}$.}
    \label{fig:number}
\end{figure}

The scaling behaviors given in Eq.~(\ref{eq:AvDistribSusp}) are confirmed by the finite-size analysis of numerical data, which consists in trying to collapse data for different sizes $L$ using appropriate $L$-dependent rescalings of both the variable and the distribution.
A good data collapse is obtained when plotting $L^{d_f \tau} P(S)$ versus $S/L^{d_f}$ (see Fig.~\ref{fig:size}), $L^{z \tau'} P(T)$ versus $T/L^{z}$ (see Fig.~\ref{fig:time}), and $L^{\chi \tau''} P(N)$ versus $N/L^{\chi}$ (see Fig.~\ref{fig:number}), for different large values of $L$. For each observable, the corresponding cutoff exponent $d_f$, $z$ or $\chi$ has been tuned to obtain the best possible collapse.
A well-defined power-law behavior is observed on the collapsed data over several decades for each of the distributions $P(S)$, $P(T)$ and $P(N)$, allowing for rather accurate estimates of the power-law exponents $\tau$, $\tau'$, and $\tau''$. Note that the scaling form with system size $L$ used for the finite-size scaling analysis implicitly assumes that $\tau$, $\tau'$, $\tau'' >1$, which is a posteriori verified numerically. Otherwise, the distributions would acquire an $L$-dependent normalization prefactor which would modify the scaling form used in the finite-size scaling analysis.

For the largest $\alpha$ value, $\alpha=3$, i.e.~in the short-range depinning limit, we observe a clear bump for large avalanches in the size and duration distributions. 
This is a known feature of depinning avalanches, also observed in one-dimensional models~\cite{laursonCriticalityInterfaceDepinning2024}.
For smaller $\alpha$ values, this bump gradually widens and flattens, as shown in Appendix~\ref{sec:app:avalanches}.

    



\subsection{Exponents evolution with $\alpha$}

\begin{table}
\caption{Measured critical exponents.}
\label{table:exponents}
\begin{ruledtabular}
\begin{tabular}{c d d d d d d} 
  \textrm{Exponent} & \multicolumn{1}{c}{$\alpha=0.5$} &
  \multicolumn{1}{c}{$\alpha=1.25$} &
  \multicolumn{1}{c}{$\alpha=1.5$} &
  \multicolumn{1}{c}{$\alpha=1.75$} &
  \multicolumn{1}{c}{$\alpha=2$} &
  \multicolumn{1}{c}{$\alpha=3$} \\
 \hline
 $d_f$   & 1.4  & 1.75 & 2    & 2.4  & 2.75 & 2.8  \\ 
 $z$      & 0.9  & 0.95 & 1.05 & 1.3  & 1.55 & 1.55 \\
 $\chi$   & 1.4  & 1.75 & 2    & 2    & 2    & 2    \\
 $\tau$   & 1.4  & 1.47 & 1.48 & 1.41 & 1.34 & 1.28 \\
 $\tau'$  & 1.6  & 1.75 & 1.8  & 1.72 & 1.57 & 1.47 \\
 $\tau''$ & 1.42 & 1.49 & 1.5  & 1.47 & 1.42 & 1.37 \\
  $\tau_\mathrm{cl}$   & 2.8  & 2.3 & 1.85 & 1.7 & 1.7 & 1.8 \\
 $\tau'_\mathrm{cl}$  &  - & - & -  & 2 & 2 & 2.2 \\
 $\tau''_\mathrm{cl}$ & 2.8 & 2.6 & 2  & 1.8 & 1.8 & 1.9 
 \label{table:exponents}
\end{tabular}
\end{ruledtabular}
\end{table}

\begin{figure}[t]
    \centering
    \includegraphics[width=0.48\textwidth]{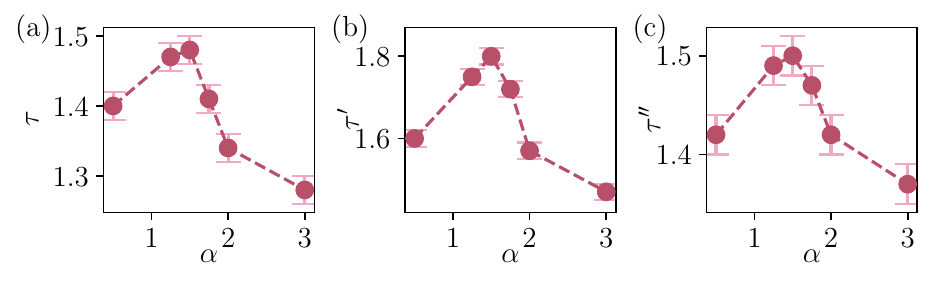}
    \includegraphics[width=0.48\textwidth]{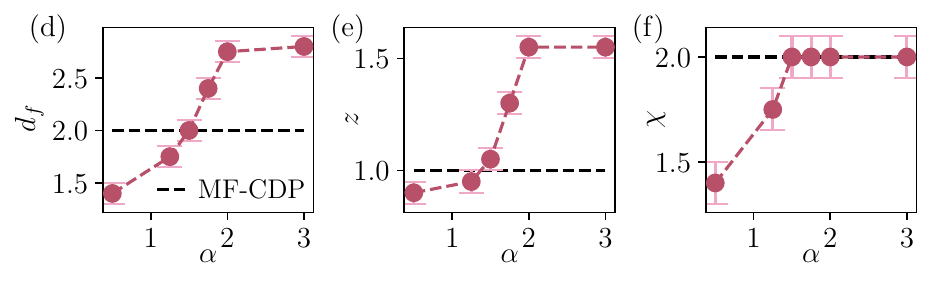}
    \caption{Evolution of the exponents characterizing the avalanche distributions $P(S)$, $P(T)$ and $P(N)$ [see Eq.~(\ref{eq:AvDistribSusp})] as a function of the decay exponent $\alpha$ of long-range interactions. Top: Power-law exponents $\tau$, $\tau'$ and $\tau''$. Bottom: Cutoff scaling exponents $d_f$, $z$ and $\chi$.}
    \label{fig:exp_evo}
\end{figure}

Following the above finite-size scaling procedure, the different exponents have been measured for all available values of the decay exponent $\alpha$ of long-range interactions, i.e.,
$\alpha \in \{ 3, 2, 1.75, 1.5, 1.25, 0.5 \}$. Exponent values are reported in Table~\ref{table:exponents}.
To better visualize the dependence of the exponents on the interaction range,
Fig.~\ref{fig:exp_evo} displays the evolution as a function of $\alpha$ of the power-law exponents $\tau$, $\tau'$, and $\tau''$ (top panels) as well as the cutoff exponents $d_f$, $z$ and $\chi$ (bottom panels).

\begin{figure}[t]
    \centering
    \includegraphics[width=\columnwidth]{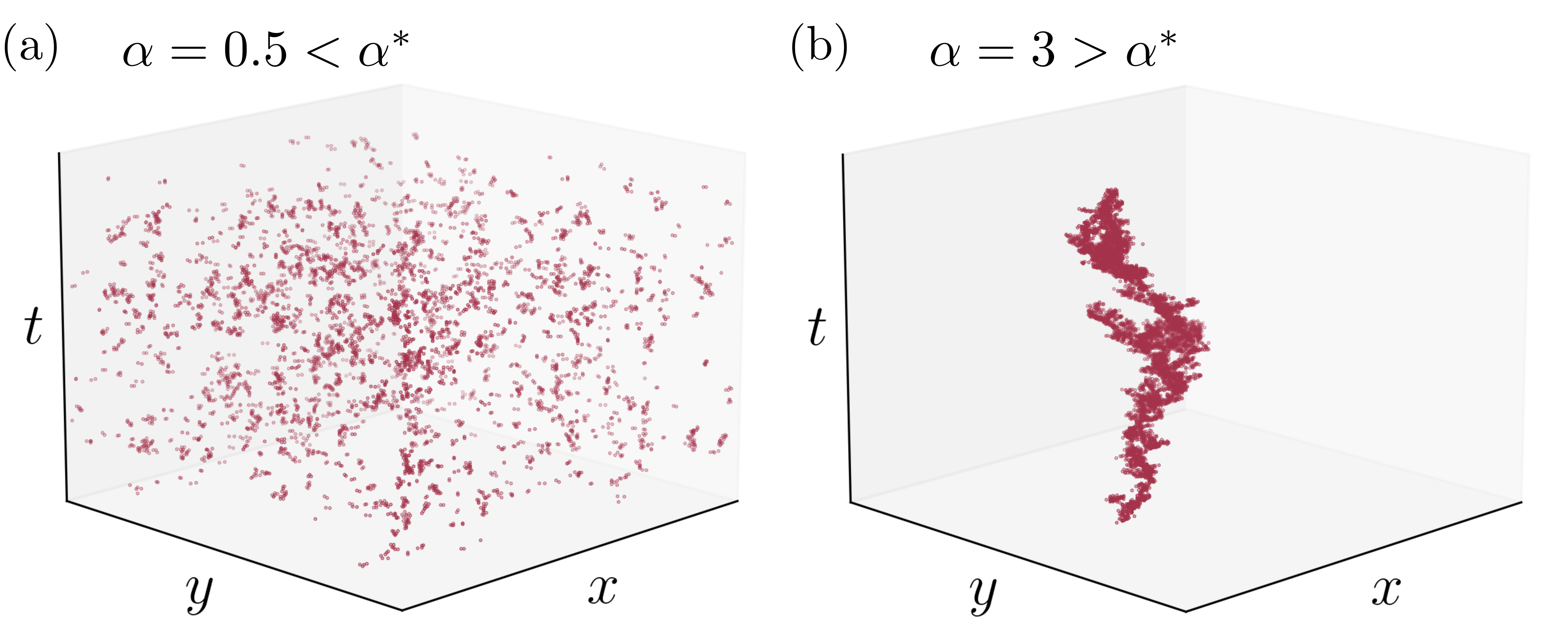}
    \caption{Space-time plots of a typical avalanche, for $\alpha=0.5$ (a), showing a non-compact avalanche, and $\alpha=3$ (b), showing a compact avalanche.}
    \label{fig:av_spacetime}
\end{figure}

A first observation is that the power-law exponents $\tau$, $\tau'$ and $\tau''$ all exhibit a qualitatively similar non-monotonous behavior. When decreasing $\alpha$ from $\alpha=3$, the values of these exponents increase from their short range value up to a maximum value, which at least for $\tau$ is compatible with the mean-field value $\tau_{\text{mf}}=1.5$, which is reached for $\alpha \approx 1.5$.
Further decreasing $\alpha$, the exponents $\tau$, $\tau'$ and $\tau''$ slightly decrease, even though their value remains significantly above their short-range value for the explored range of $\alpha$.

The behavior of the cutoff exponents is even more interesting, and might shed some light on the global $\alpha$-dependence of the critical behavior.
Starting from low values of $\alpha$, the three cutoff exponents monotonously increase before reaching a plateau value. 
The behavior of the fractal dimension $d_f$ of avalanches is particularly intriguing, as it crosses the space dimension $d=2$ for $\alpha = \alpha^* \approx 1.5$.
Hence for $\alpha < \alpha^*$, one has $d_f <d$ so that avalanches are non-compact. Conversely, for $\alpha > \alpha^*$, $d_f >d$ meaning that avalanches are compact and that the avalanche size is much larger than the number of particles (or the system volume). This means that active particles are activated a large number of times during an avalanche.
This behavior is illustrated in Fig.~\ref{fig:av_spacetime}, where we plot for $\alpha < \alpha^*$ and  $\alpha > \alpha^*$ the space-time activity of two avalanches of comparable sizes within the power law regime of $P(S)$.
Note that the value $d_f=d$ also corresponds to the mean-field prediction for $d_f$ for the conserved directed percolation class~\cite{wieseTheoryExperimentsDisordered2022}.
The cutoff exponent $\chi$ characterizing the maximum  number of particles involved in an avalanche displays a similar behavior to $d_f$ for $\alpha < \alpha^*$, but saturates to $\chi=2$ for $\alpha > \alpha^*$
since  the largest avalanches cannot involve numbers of particles that grow faster than the system volume $L^d$.
Finally, the dynamical exponent $z$ also crosses the mean-field CDP value $z_{\text{mf}}=d/2$ \cite{lubeckUniversalScalingBehavior2004} for $\alpha$ in the range $1<\alpha<2$, but the actual crossing value of $\alpha$ seems to be around $\alpha \approx 1.3$, although a crossing at $\alpha \approx 1.5$ cannot be fully discarded.

Altogether, these results suggest that the non-monotonic behavior of the power-law avalanche exponents $\tau$, $\tau'$ and $\tau''$ as a function of $\alpha$ may be related to a change of geometric nature of avalanches, from compact to non-compact when decreasing $\alpha$ upon crossing the value $\alpha = \alpha^* \approx 1.5$.
The compact regime observed for $\alpha > \alpha^*$ is qualitatively similar to the long-range CDP (LR-CDP) case, for which one has in particular $d_f\geq d$, $z \geq d/2$ and $\chi=d$~\cite{wieseTheoryExperimentsDisordered2022}.
In contrast, the exponent values reported in the non-compact regime $\alpha < \alpha^*$ are not compatible with the LR-CDP scenario.
This result is consistent with recent results on  static critical exponents in the $\alpha$-ROM model, which also escape the LR-CDP range of static exponent values when $\alpha < \alpha^* \approx 1.5$
in two dimensions \cite{jocteur2025EPJE}.
This regime has been shown to be dominated by the anomalous diffusion of passive particles induced by mediated long-range interactions with distant active particles \cite{jocteur2025EPJE}.


\subsection{Mean observables conditioned to avalanche size}

\begin{figure}[t]
    \centering
    \includegraphics[width=\columnwidth]{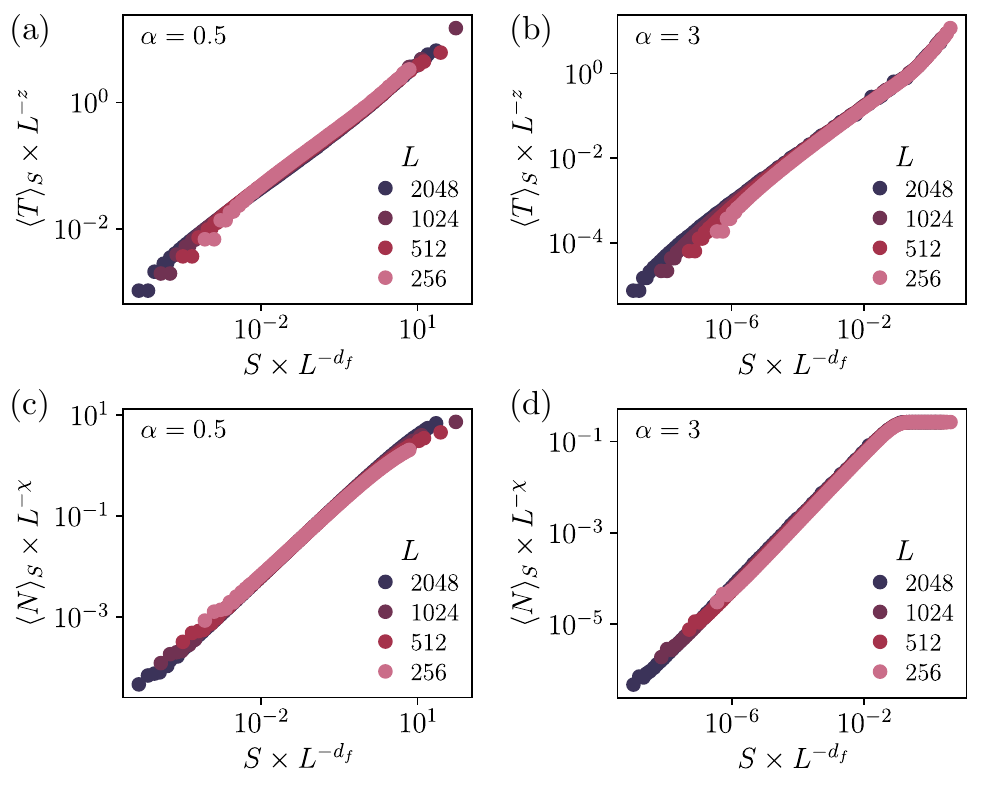}
    \caption{(a) and (b) Finite-size scaling analysis of the relation between avalanche size $S$ and average duration $\langle T\rangle_S$ of avalanches of size $S$, for $\alpha=0.5$ (a) and $\alpha=3$ (b).
    (c) and (d) Finite-size scaling analysis of the relation between avalanche size $S$ and average  number of particles involved $\langle N\rangle_S$, for $\alpha=0.5$ (c) and $\alpha=3$ (d).}
    \label{fig:ST_SN}
\end{figure}

We have measured above the distributions of avalanche size, duration and number of involved particles independently one from the other. It is also of interest to quantify how these quantities correlate one with another. For instance, it may be expected that an avalanche with a large (resp.~small) size also has a large (resp.~small) duration. For broadly distributed variables, directly computing correlations (i.e., covariances) may not be convenient, since averages are typically dominated by the largest values, so that the correlation between small events may hardly be visible.
We thus rather evaluate the average values $\langle T\rangle_S$ and $\langle N\rangle_S$ conditioned to a fixed avalanche size $S$, allowing us to characterize correlations between observables in both small and large avalanches.

In Fig.~\ref{fig:ST_SN} we plot the resulting data in terms of a finite-size scaling analysis, using the natural scaling variables $S/L^{d_f}$, $\langle T\rangle_S/L^z$ and $\langle N\rangle_S/L^{\chi}$.
Panels (a) and (b) display the rescaled conditional average duration $\langle T\rangle_S/L^z$ versus rescaled avalanche size $S/L^{d_f}$, for $\alpha=0.5$ and $\alpha=3$ respectively.
Similarly, panels (c) and (d) report the rescaled conditional average number of involved particles $\langle N\rangle_S/L^{\chi}$ versus rescaled avalanche size $S/L^{d_f}$.
A good data collapse is observed for both observables.
This means that all avalanches, irrespective of their size, have statistically the same spatial and temporal structure, governed by the  cutoff exponents that we determined for the largest avalanches.
Additional plots for intermediate values of $\alpha$ are reported in Appendix~\ref{sec:app:avalanches} (see Figs.~\ref{fig:ST:app} and \ref{fig:SN:app}).




\subsection{Scaling relations}
\begin{figure}[t]
    \centering
    \includegraphics[width=0.6\columnwidth]{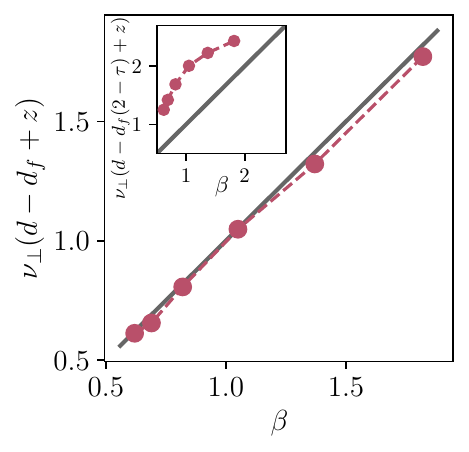}
    \caption{Test of the scaling relation, Eq.~(\ref{eq:scaling_flow}): Exponent combination $\nu_\perp (d-d_f+z)$ versus $\beta$ (red symbols). The grey line corresponds to the equality $\nu_\perp (d-d_f+z)=\beta$ [i.e., Eq.~(\ref{eq:scaling_flow})]. Inset: test of the modified relation $\nu_\perp (d-d_f(2-\tau)+z)=\beta$.}
    \label{fig:scaling_flow}
\end{figure}

We now test the usual depinning-like scaling picture for which the steady dynamics in the active phase just above the transition ($\phi >\phi_\mathrm{c}$) is well approximated as a succession of avalanches of linear size of the order of the correlation length $\xi$~\cite{lin_scaling_2014}.
This picture leads to the scaling relation
\begin{equation}
    \beta = \nu_\perp (d-d_f+z)\, ,
    \label{eq:scaling_flow}
\end{equation}
with $\nu_\perp$ the exponent of the correlation length $\xi$, such that $\xi\sim (\phi-\phi_\mathrm{c})^{-\nu\perp}$.
We estimate the value of $\nu_\perp$ assuming hyperscaling $\nu_\perp = d^{-1}(2\beta + \gamma^\prime)$, where $\beta$ and $\gamma^\prime$ are the order parameter and order parameter fluctuations exponents, respectively, the values of which were numerically evaluated in~\cite{jocteur2025EPJE}.
With this estimate, we test Eq.~(\ref{eq:scaling_flow}) in Fig.~\ref{fig:scaling_flow} by plotting $\nu_\perp (d-d_f+z)$ versus $\beta$ for all the $\alpha$ values we investigated.
The grey line corresponds to Eq.~(\ref{eq:scaling_flow}), which is very well followed by the numerical estimates of the exponents.
We note that a modified scaling relation $\beta = \nu_\perp(d-d_f(2-\tau) + z)$ was advocated in the context of yielding, based on molecular dynamics results~\cite{relmucao-leivaBridgingGapAvalanche2025}.
The difference with Eq.~(\ref{eq:scaling_flow}) is whether the dynamics is dominated by the largest avalanches as in Eq.~(\ref{eq:scaling_flow}), or by mean avalanches, leading to the modified scaling.
Based on the facts that (i) Eq.~(\ref{eq:scaling_flow}) seems to hold for our data, and (ii) our measured $\tau$ values are significantly larger than $1$, we can anticipate that the modified scaling does not hold in our case, as confirmed by data plotted in inset of Fig.~\ref{fig:scaling_flow}.

The fact that Eq.~(\ref{eq:scaling_flow}) holds is both a test of the depinning-like scaling picture and a hint that  $d_f-z$ does not depend on the reactivation protocol. This is important as it may be that $d_f$ and $z$ taken separately do depend on the way we reactivate the dynamics, by analogy with the yielding transition~\cite{jocteurProtocol2025}.

Finally, in the context of yielding, another scaling relation involving $\nu_\perp$ and $d_f$ has been proposed, $\nu_\perp^{-1} = \alpha-d_f$, based this time on the assumption of statistical tilt symmetry~\cite{lin_scaling_2014}.
In our case, however, the exponent $\alpha$ is associated with a long-range interaction -- long-range mechanical noise -- of a different nature than in the case of yielding, where $\alpha$ is associated with long-range transport of the conserved quantity. 
We thus do not expect the scaling relation to hold for the $\alpha$-ROM, and indeed it is obvious that it does not for the small values of $\alpha$, for which $d_f>\alpha$ (see Fig.~\ref{fig:exp_evo}),
so that the scaling relation $\nu_\perp^{-1} = \alpha-d_f$ is incompatible with the basic expectation of a positive $\nu_\perp$ (i.e., divergence of the correlation length at the critical point).

\subsection{Clusters}

\begin{figure}[t]
    \centering
    \includegraphics[width=\columnwidth]{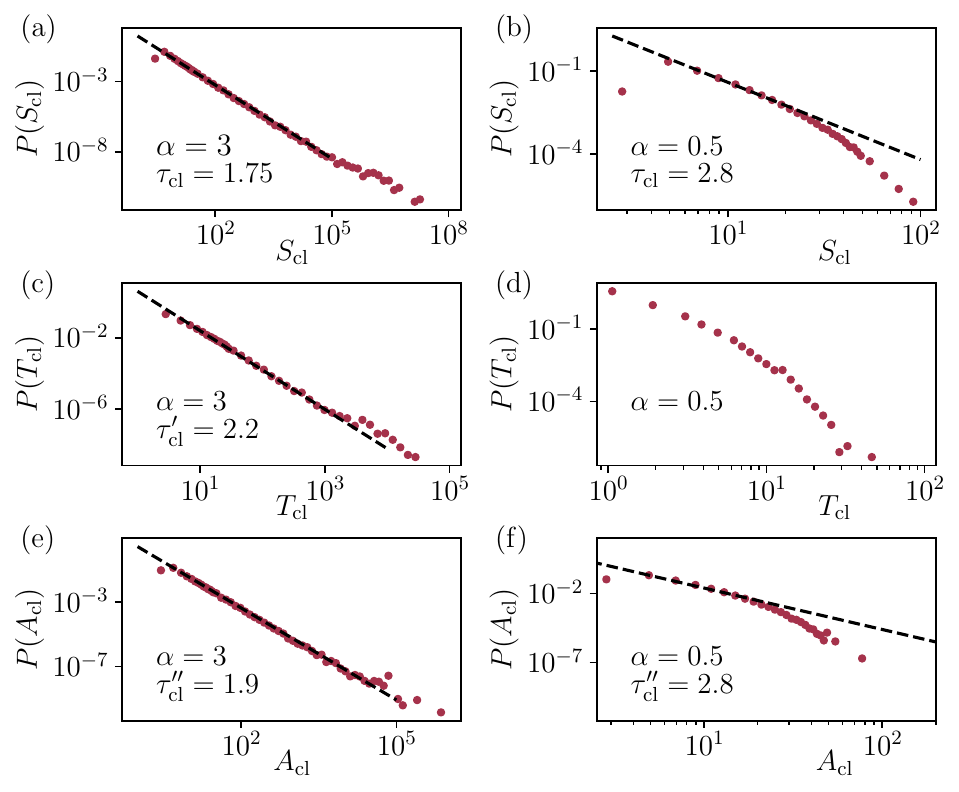}
    \caption{Cluster statistics, for $\alpha=3$ (panels a, c, e) and $\alpha=0.5$ (panels b, d, f). (a) and (b): Cluster size distribution. (c) and (d): Cluster duration distribution. (e) and (f): Cluster area distribution. In dashed lines, we show power law fits to the distributions, except in (d) where the distribution does not show any range over which a power law is a reasonable fit.}
    \label{fig:cluster_distributions}
\end{figure}

Avalanches in systems with long-ranged transport are formed of spatially disconnected clusters~ \cite{maloyLocalWaitingTime2006,laursonAvalanchesClustersPlanar2010a,planetSpatiotemporalOrganizationCorrelated2018,lepriolSpatialClusteringDepinning2021,caoClustersEpidemicModel2022},
whose properties (size, spatial extension, duration) are also power-law distributed. 
In the context of long-range depinning, the cluster statistics is different from the one of full avalanches but related to it through scaling relations ~\cite{laursonAvalanchesClustersPlanar2010a,lepriolSpatialClusteringDepinning2021}, which makes them useful observables to indirectly characterize avalanches.

We thus characterize the cluster statistics for the $\alpha$-ROM, as a function of the range exponent $\alpha$.
As the $\alpha$-ROM is defined in a continuous space, in order to define clusters, we spatially discretize particle activity on a regular two-dimensional grid with spacing taken equal to the particle diameter $D=1$.
A given grid cell at position $(i,j)$ is declared active at time $t$ if the center of a particle active at time $t$ is within the cell.
If we associate with each active cell in an avalanche a vertex $(i,j,t)$ we can build a graph with edges connecting vertices $(i,j,t)$ and $(i',j',t')$ such that $\sqrt{(i-i')^2 + (j-j')^2 + (t-t')^2}\leq 2\sqrt{3}$. 
This graph encodes the spatio-temporal structure of the avalanche.
Clusters are then simply defined as the connected components of the graph. They correspond to bursts of activity that occur contiguously in space and time.

Each cluster has a size $S_\mathrm{cl}$ which is the total number of vertices in the connected component, a duration $T_\mathrm{cl}$ defined as the difference between smallest and largest time in the cluster, and an area $A_\mathrm{cl}$ defined as the number of different cells
in the cluster. Note that these definitions parallel the ones of the size $S$, duration $T$ and particle number $N$ of an avalanche, but are slightly different due to the spatial discretization of activity onto a grid.
In particular, because several particles can be active in a cell, the size of an avalanche can be strictly larger than the sum of the sizes of the clusters that make it up.

Based on observations on depinning and epidemic models~ \cite{maloyLocalWaitingTime2006,laursonAvalanchesClustersPlanar2010a,planetSpatiotemporalOrganizationCorrelated2018,lepriolSpatialClusteringDepinning2021,caoClustersEpidemicModel2022}, we anticipate that the distribution of these cluster characteristics follow power laws $P(S_\mathrm{cl})\sim S_\mathrm{cl}^{-\tau_\mathrm{cl}}$, $P(S_\mathrm{cl})\sim T_\mathrm{cl}^{-\tau'_\mathrm{cl}}$ and $P(S_\mathrm{cl})\sim A_\mathrm{cl}^{-\tau''_\mathrm{cl}}$ on intermediate ranges of $S_\mathrm{cl}$, $T_\mathrm{cl}$ and $A_\mathrm{cl}$.
This is indeed what we observe for the $\alpha$-ROM, as shown in Fig.~\ref{fig:cluster_distributions} for the limiting case $\alpha=3$ (panels (a), (c) and (e)). 
Much like for the avalanche statistics, a bump is observed for the largest avalanches in all cluster distributions.
By contrast, for the long-range limiting case $\alpha=0.5$, we observe only a narrow power-law range for the size and area distributions, without bump (panels (b) and (f)), and no power law at all for the duration distribution (panel (d)).
For intermediate values of $\alpha$, shown in Appendix~\ref{sec:app:clusters} (see Figs.~\ref{fig:cluster_distributions_tauc:app}, \ref{fig:cluster_distributions_taucp:app} and \ref{fig:cluster_distributions_taucpp:app}), we observe that the power-law ranges in the distributions gradually disappear when decreasing $\alpha$ between $\alpha=3$ and $\alpha=0.5$.

\begin{figure}[t]
    \centering
    \includegraphics[width=0.7\columnwidth]{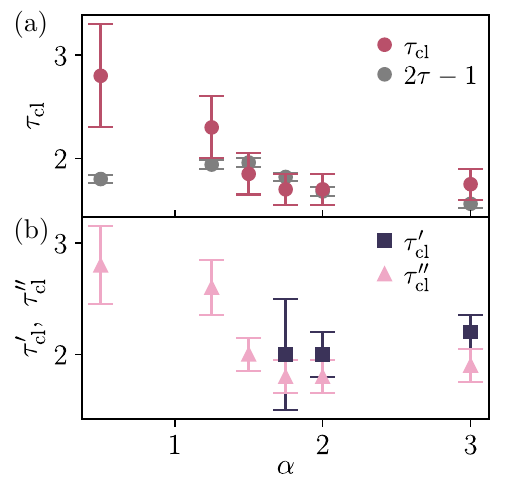}
    \caption{Cluster exponents, as a function of the range exponent $\alpha$. (a) Size exponent $\tau_\mathrm{cl}$ (dark red symbols). In grey symbols, we show for comparison $2\tau-1$. (b) Duration exponent $\tau'_\mathrm{cl}$ (dark blue) and area exponent $\tau''_\mathrm{cl}$ (pink).}
    \label{fig:cluster_exponents}
\end{figure}

The measured exponents $\tau_\mathrm{cl}$, $\tau'_\mathrm{cl}$, $\tau''_\mathrm{cl}$ are shown as a function of $\alpha$ in Fig.~\ref{fig:cluster_exponents}. 
Generically, all these exponents have values which increase when the interaction range increases (i.e., $\alpha$ decreases), which was also observed in one-dimensional long-range depinning~\cite{lepriolSpatialClusteringDepinning2021}.
In the latter context, the exponent $\tau_\mathrm{cl}$ was argued to be related to the avalanche size exponent $\tau$ via the scaling relation $\tau_\mathrm{cl} = 2\tau - 1$~\cite{laursonAvalanchesClustersPlanar2010a,lepriolSpatialClusteringDepinning2021}.
We observe that this scaling relation is compatible with our measurements in the $\alpha$-ROM for $\alpha\geq 1.5$ (in particular in the depinning limit $\alpha=3$), but not for smaller values of $\alpha$, as shown in Fig.~\ref{fig:cluster_exponents}(a) where we represented $2\tau-1$ (grey symbols) alongside $\tau_\mathrm{cl}$ (dark blue).


\section{Discussion}

In this work, we have investigated avalanche dynamics at the absorbing phase transition of a mediated Random Organization Model, designed to capture the long-range displacements induced by hydrodynamic interactions in cyclically sheared suspensions. By introducing a power-law kernel $G(r)\sim r^{-2\alpha}$ and varying the decay exponent $\alpha$, we continuously interpolated between an effectively short-range ROM and a strongly long-range, mean-field-like regime, while keeping a well-defined absorbing phase transition at $\phi=\phi_\mathrm{c}(\alpha)$.

At criticality, we have shown that the dynamics is organized into scale-free avalanches. The distributions of avalanche size $P(S)$, duration $P(T)$, and number of involved particles $P(N)$ follow robust power laws over several decades with system-size cutoffs governed by exponents $d_f$, $z$, and $\chi$, respectively. Crucially, all these exponents evolve with $\alpha$, demonstrating that the interaction range directly controls avalanche statistics and geometry at the absorbing transition.

A central result is the qualitative change in avalanche spatial structure upon varying $\alpha$. The fractal dimension $d_f$ crosses the space dimension $d=2$ around $\alpha^*\approx 1.5$. For $\alpha<\alpha^*$, avalanches are non-compact, in the sense that $d_f <d$. In contrast, for $\alpha>\alpha^*$, avalanches become effectively compact ($d_f>d$), implying that particles are activated many times within a single avalanche, while the scaling exponent of the number of distinct participating particles saturates to the trivial bound $\chi=d$ set by the system volume. This crossover may provide a tentative geometrical interpretation of the non-monotonic behavior of the power-law exponents $\tau$, $\tau'$ and $\tau''$ observed when scanning $\alpha$.

Beyond the $\alpha$-dependence of individual exponents, our results also  validate the depinning-like scaling picture based on the dominance of large avalanches in the global activity in the active phase near the transition  in the context of the RIT, for all the $\alpha$ values we investigated.

Overall, our study establishes avalanche statistics as a sensitive probe of interaction-range effects at nonequilibrium absorbing phase transitions. The $\alpha$-ROM provides a minimal setting where hydrodynamic-like long-range couplings can be tuned and their impact on avalanche geometry and scaling relations can be cleanly assessed. 
This opens several natural perspectives, in particular regarding a systematic characterization of the internal structure of avalanche clusters (using, e.g., space and time correlation functions), and 
 a more detailed test of scaling scenarios in the crossover region near $\alpha^*$, where compactness changes.
Finally, our results should be tested by direct comparison with experiments on oscillatory sheared suspensions, where long-range fluid-mediated couplings are unavoidable, and where reactivation could be provided by Brownian motion~\cite{keimGenericTransientMemory2011,paulsenMultipleTransientMemories2014} or sedimentation~\cite{corteSelfOrganizedCriticalitySheared2009}.

\acknowledgments
This project was provided with computer and storage resources by GENCI at
IDRIS thanks to the grants 2024-AD010914551R1 and 2025-AD010914551R2
on the supercomputer Jean Zay's V100 and A100 partitions. 
Some of the computations presented in this paper were performed using the GRICAD infrastructure (\href{https://gricad.univ-grenoble-alpes.fr}{https://gricad.univ-grenoble-alpes.fr}), which is supported by Grenoble research communities.
T.J. acknowledges funding from the French Ministry of Higher Education and Research.

\section*{Author contributions}

T.J., K.M., E.B. and R.M. designed the study. T.J. and R.M. developed the numerical code. T.J. performed the simulations.
T.J. and R.M. performed the data analysis. K.M., E.B. and R.M. supervised the project. T.J., K.M., E.B. and R.M. wrote the manuscript.

\section*{Data availability}

Data sets generated during the current study are available from the corresponding author upon reasonable request.


\appendix

\section{Avalanche distributions for all simulated $\alpha$ values}
\label{sec:app:avalanches}

\begin{figure}[t]
    \centering
    \includegraphics[width=0.34\textwidth]{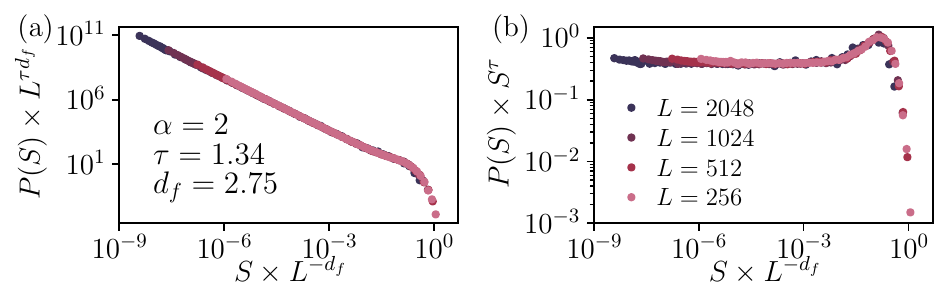}
    \includegraphics[width=0.34\textwidth]{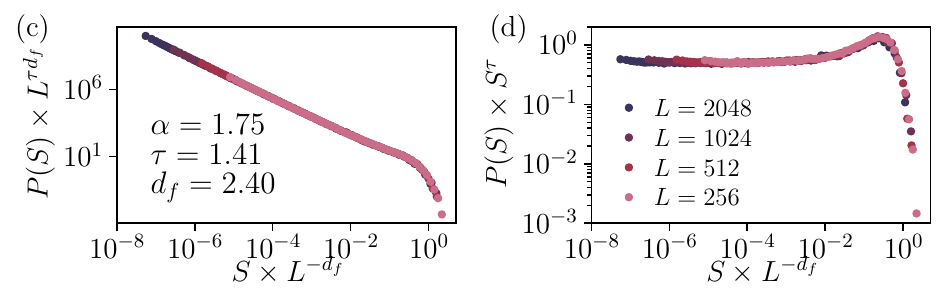}
    \includegraphics[width=0.34\textwidth]{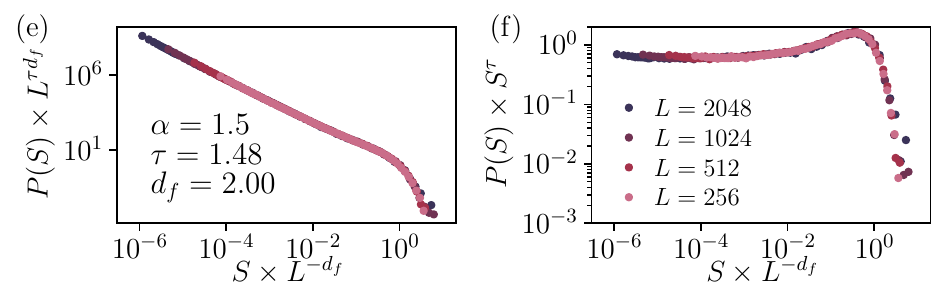}
    \includegraphics[width=0.34\textwidth]{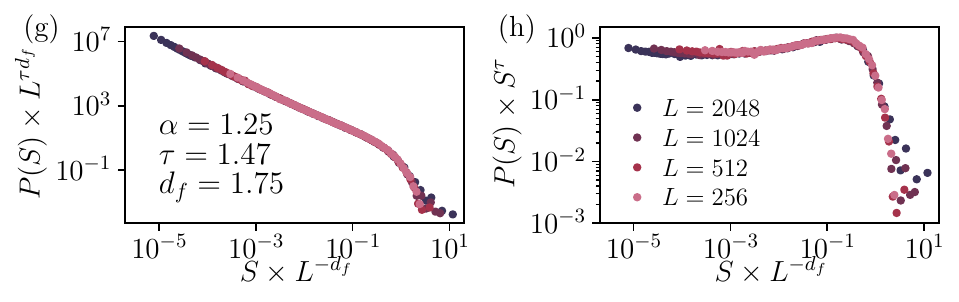}
    \caption{Finite-size scaling of the avalanche size distributions $P(S)$ for $\alpha \in \{ 2, 1.75, 1.5, 1.25 \}$ (top to bottom). Left panels: collapse of the curves obtained by finite-size scaling. Right panels: compensated plots using the estimated power-law exponent.}
    \label{fig:size:app}
\end{figure}
\begin{figure}[t]
    \centering
    \includegraphics[width=0.34\textwidth]{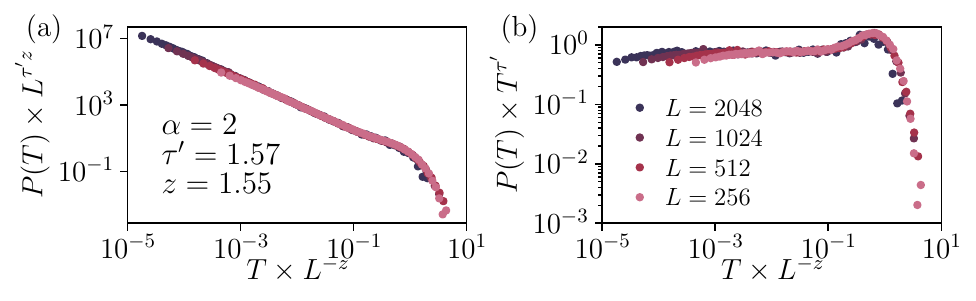}
    \includegraphics[width=0.34\textwidth]{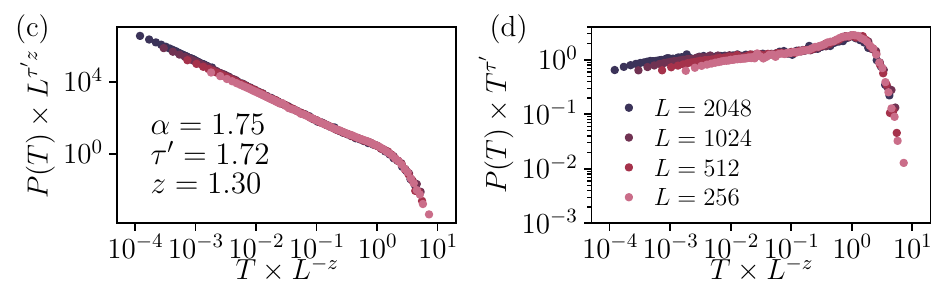}
    \includegraphics[width=0.34\textwidth]{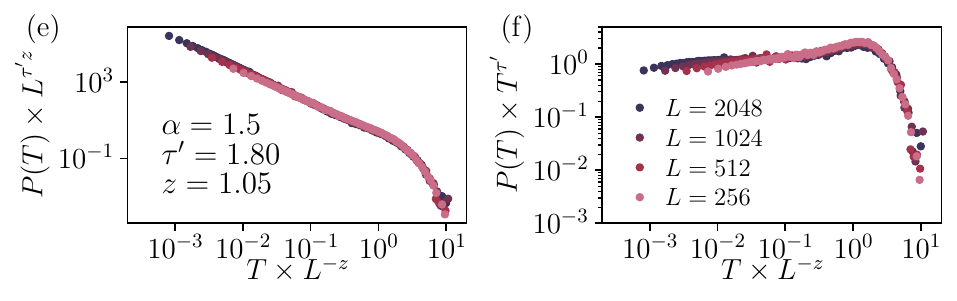}
    \includegraphics[width=0.34\textwidth]{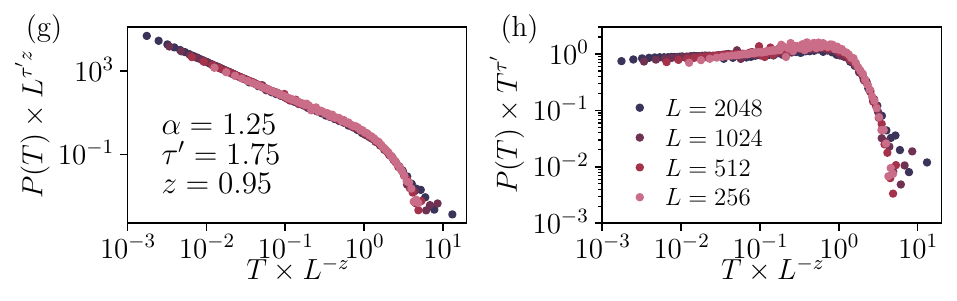}
    \caption{Finite-size scaling of the avalanche duration distributions $P(T)$ for $\alpha \in \{ 2, 1.75, 1.5, 1.25 \}$ (top to bottom). Left panels: collapse of the curves obtained by finite-size scaling. Right panels: compensated plots using the estimated power-law exponent.}
    \label{fig:time:app}
\end{figure}
\begin{figure}[t]
    \centering
    \includegraphics[width=0.34\textwidth]{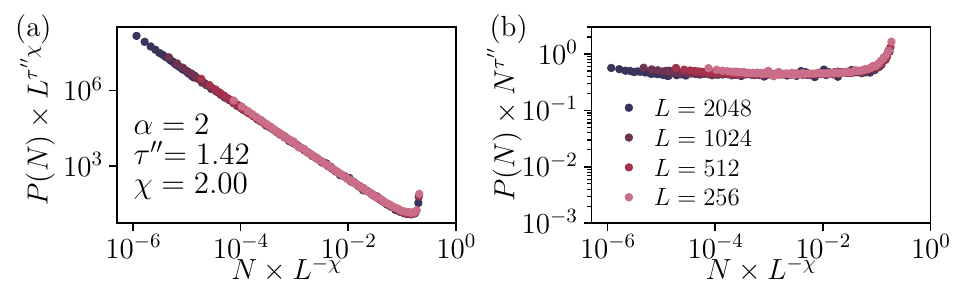}
    \includegraphics[width=0.34\textwidth]{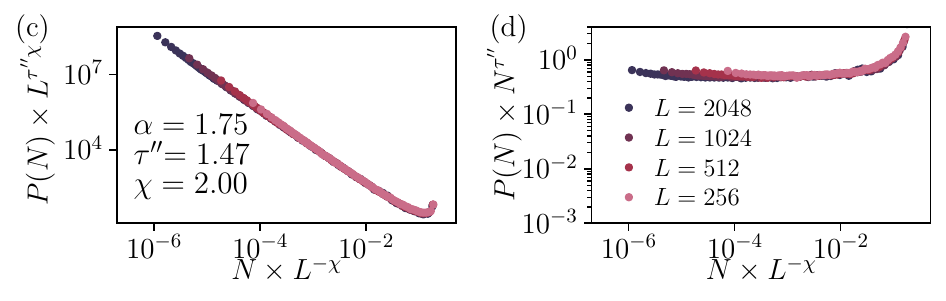}
    \includegraphics[width=0.34\textwidth]{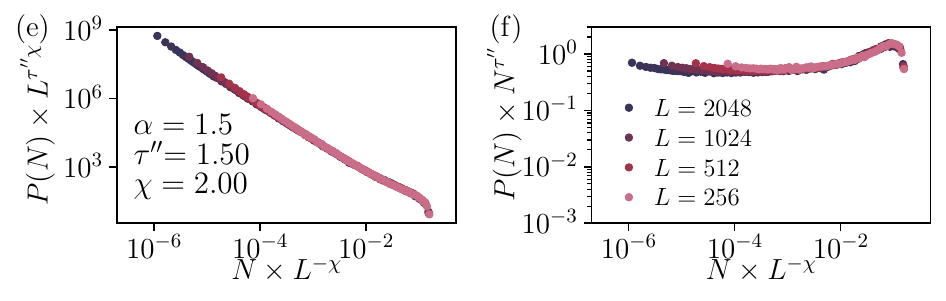}
    \includegraphics[width=0.34\textwidth]{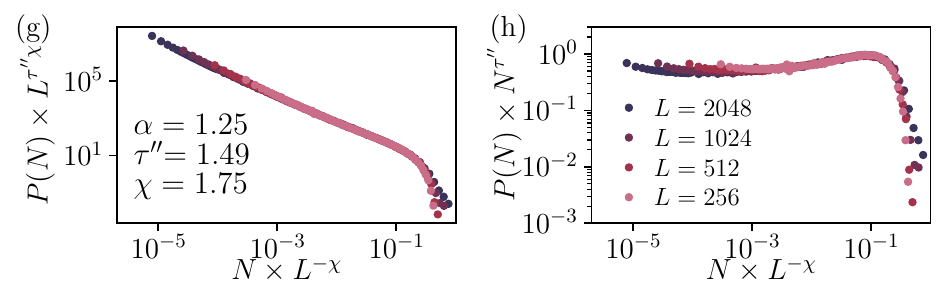}
    \caption{Finite-size scaling of the distributions $P(N)$ of the number $N$ of particles involved in an avalanche, for $\alpha \in \{ 2, 1.75, 1.5, 1.25 \}$ (top to bottom). Left panels: collapse of the curves obtained by finite-size scaling. Right panels: compensated plots using the estimated power-law exponent.}
    \label{fig:number:app}
\end{figure}

In the main text, we displayed the distributions of several avalanche observables (size $S$, duration $T$ and number $N$ of particles involved), as well as the relation between average values of these observables, focusing on two well-separated values of the power-law decay exponent $\alpha$ of long-range interactions, $\alpha=3$ and $\alpha=0.5$. For the sake of completeness, we report here the same observables for several intermediate values of $\alpha$.
The distributions of avalanche size $S$, duration $T$ and number $N$ of involved particles are shown in Figs.~\ref{fig:size:app}, \ref{fig:time:app} and \ref{fig:number:app} respectively,
for $\alpha \in \{ 2, 1.75, 1.5, 1.25 \}$ and system size $L \in \{ 256, 512, 1024, 2048\}$.

This shows the continuous evolution of the avalanche exponents with $\alpha$. Also, we observe the large avalanche bump gradually spreading when $\alpha$ decreases, i.e. going to longer ranged interactions.
\begin{figure}[t]
    \centering
    \includegraphics[width=0.8\columnwidth]{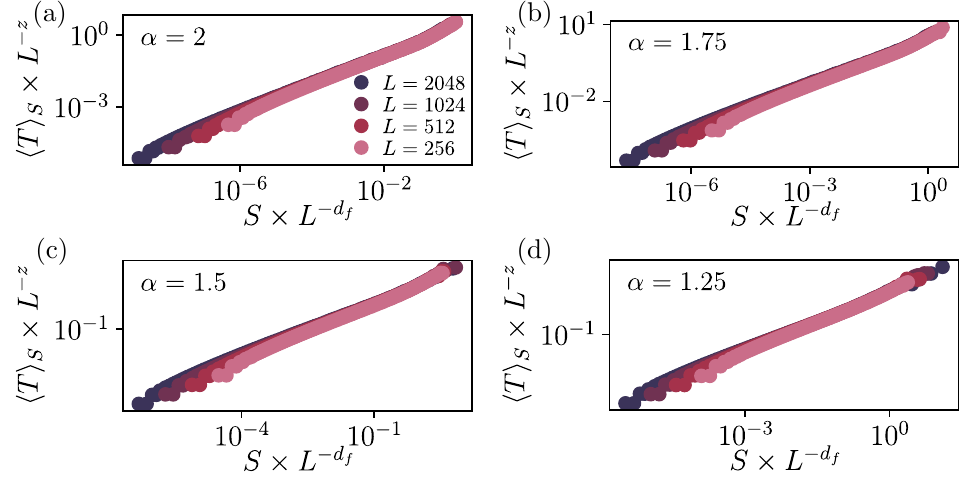}
    \caption{Finite-size scaling analysis of the relation between duration $\langle T\rangle_S$ and  avalanche size $S$, for $\alpha \in \{ 2, 1.75, 1.5, 1.25 \}$.}
    \label{fig:ST:app}
\end{figure}

\begin{figure}[t]
    \centering
    \includegraphics[width=0.8\columnwidth]{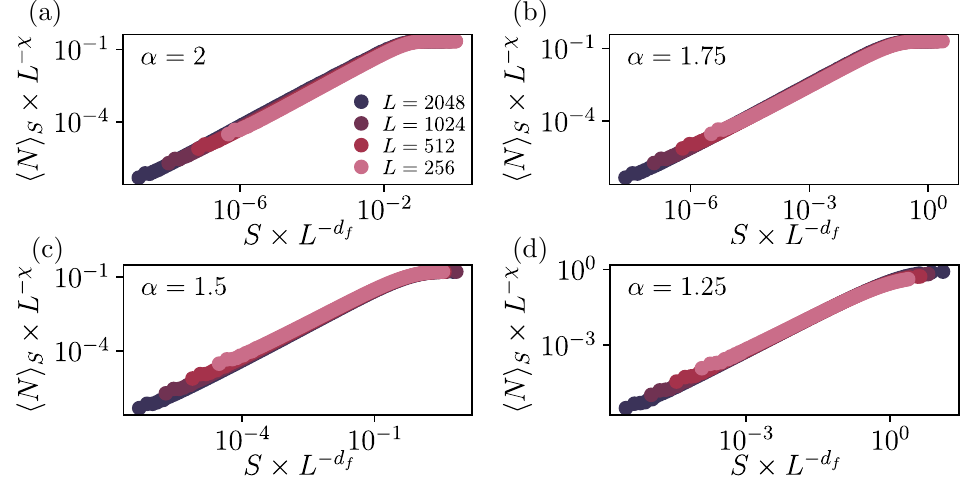}
    \caption{Finite-size scaling analysis of the relation between average number $\langle N\rangle_S$ of particles involved in an avalanche and  avalanche size $S$, for $\alpha \in \{ 2, 1.75, 1.5, 1.25 \}$.}
    \label{fig:SN:app}
\end{figure}

Figs.~\ref{fig:ST:app} and \ref{fig:SN:app} further display the rescaled average avalanche duration $\langle T\rangle_S$ [Fig.~\ref{fig:ST:app}] and
average number $\langle N\rangle_S$ of particles involved in an avalanche [Fig.~\ref{fig:SN:app}], conditioned to a given avalanche size $S$, for the same sets of values of $\alpha$ and $L$.

\section{Cluster distributions for all simulated $\alpha$ values}
\label{sec:app:clusters}

\begin{figure}[t]
    \centering
    \includegraphics[width=0.8\columnwidth]{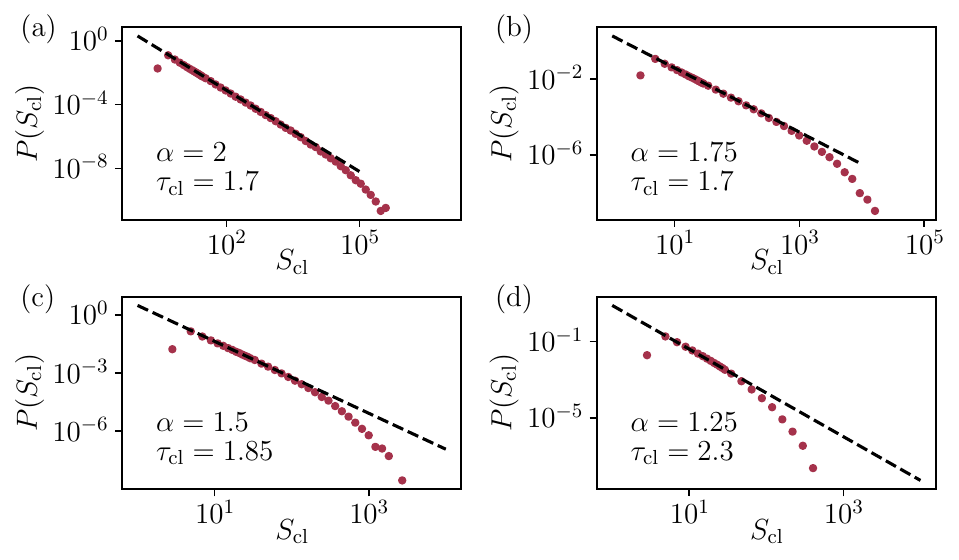}
    \caption{Cluster size distributions, for $\alpha \in \{ 2, 1.75, 1.5, 1.25 \}$. In dashed lines, power-law fits with exponents $\tau_\mathrm{cl}$ indicated on each panel.}
    \label{fig:cluster_distributions_tauc:app}
\end{figure}

\begin{figure}[t]
    \centering
    \includegraphics[width=0.8\columnwidth]{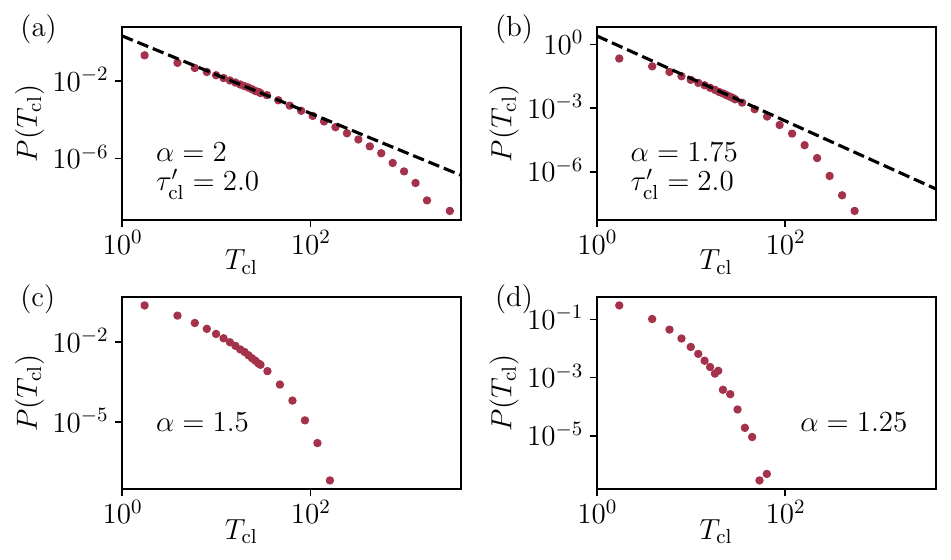}
    \caption{Cluster duration distributions, for $\alpha \in \{ 2, 1.75, 1.5, 1.25 \}$. In dashed lines, power-law fits with exponents $\tau'_\mathrm{cl}$ indicated for panels (a) and (b).}
    \label{fig:cluster_distributions_taucp:app}
\end{figure}

\begin{figure}[t]
    \centering
    \includegraphics[width=0.8\columnwidth]{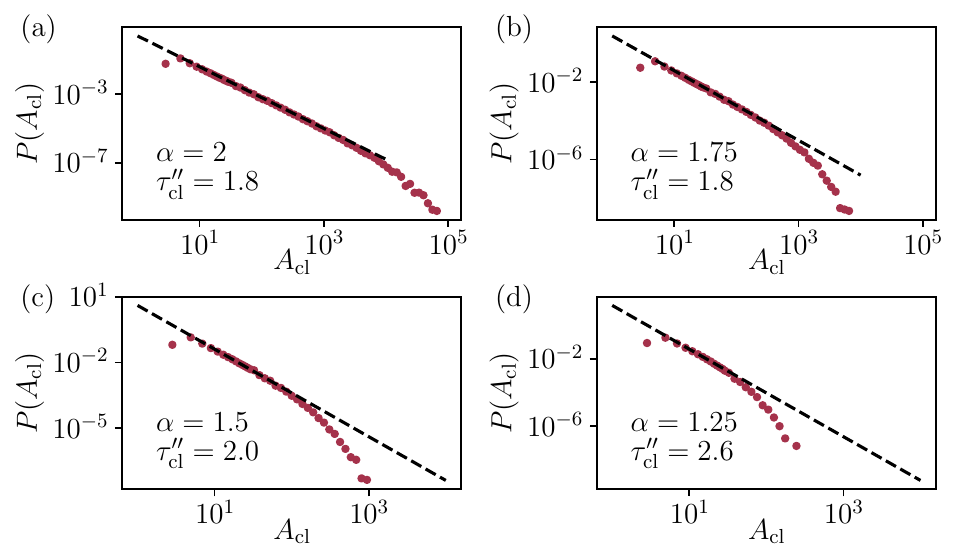}
    \caption{Cluster area distributions, for $\alpha \in \{ 2, 1.75, 1.5, 1.25 \}$. In dashed lines, power-law fits with exponents $\tau''_\mathrm{cl}$ indicated on each panel.}
    \label{fig:cluster_distributions_taucpp:app}
\end{figure}

In this appendix, we report the distributions of cluster characteristics for intermediate values of $\alpha$ (as compared to the main text), $\alpha \in \{ 2, 1.75, 1.5, 1.25 \}$.
Figs.~\ref{fig:cluster_distributions_tauc:app}, \ref{fig:cluster_distributions_taucp:app} and \ref{fig:cluster_distributions_taucpp:app} respectively display the distributions
of cluster size $S_\mathrm{cl}$, duration $T_\mathrm{cl}$ and area $A_\mathrm{cl}$.

\newpage

\bibliographystyle{apsrev4-2}
\bibliography{biblio}

\end{document}